\documentclass[journal,article,submit,moreauthors,pdftex]{Definitions/mdpi} 

\firstpage{1} 
\pubvolume{1}
\issuenum{1}
\articlenumber{0}
\pubyear{2021}
\copyrightyear{2020}
\datereceived{} 
\dateaccepted{} 
\datepublished{} 
\hreflink{https://doi.org/} 

\Title{Special Issue on Advances in Chiral Quark Models}

\TitleCitation{Special Issue on Advances in Chiral Quark Models}

\Author{Jorge Segovia$^{1}$\orcidA{}}

\AuthorNames{Jorge Segovia}

\AuthorCitation{Segovia, J.}

\address{%
$^{1}$ \quad Departamento de Sistemas F\'isicos, Qu\'imicos y Naturales, Universidad Pablo de Olavide, E-41013 Sevilla, Spain; jsegovia@upo.es\\
}

\abstract{
The number of exotic candidates in both light- and heavy-quark hadron sectors has increased dramatically since the discovery by the Belle Collaboration of the so-called $X(3872)$ in 2003. It is clear that the simple quark model picture needs an extension and thus the last twenty years have witnessed an explosion of related theoretical and experimental activity.
The ultimate goal of theory is to describe the properties of exotic states from the first principles of Quantum Chromodynamics (QCD), which is the non-Abelian Quantum Field Theory that describes the strong interaction. However, since this task is quite challenging, a more modest goal to start with is the development of QCD-motivated phenomenological models that specify the colored constituents, how they are clustered, and the forces between them. 
This Special Issue invited contributions reporting recent advances of phenomenological quark models in the study of hadron's spectrocopy, structure, and interactions, paying special attention to the exotic candidates but without losing sight of the conventional states. 
In response to the call for papers, and after a comprehensive peer review process, 8 articles qualified for acceptance in the final edition of the Special Issue. The authors are from geographically distributed countries such as Spain, South Africa, Ghana, China, Brazil, Argentina. This reflects the impact of the proposed topic and the effective organization of the guest editorial team of this Special Issue.
}

\begin{document}

\section{Introduction}

QCD is the strong-interaction part of the Standard Model of Particle Physics and solving it presents a fundamental problem that is unique in the history of science. Never before have we been confronted by a theory whose elementary excitations (quarks and gluons) are not those degrees-of-freedom readily accessible via experiment; \emph{i.e.}, whose elementary excitations are confined. Moreover, there are numerous reasons to believe that QCD generates forces which are so strong that less-than 2\% of a nucleon's mass can be attributed to the so-called current-quark masses that appear in QCD's Lagrangian; \emph{viz.}, forces capable of generating mass from nothing, a phenomenon known as dynamical chiral symmetry breaking (DCSB).

Neither confinement nor DCSB is apparent in QCD's Lagrangian and yet they play the dominant role in determining the observable characteristics of hadrons. This complexity makes hadron spectroscopy, the collection of readily accessible states constituted from gluons and quarks, the starting point for all further investigations. A very successful classification scheme for hadrons in terms of their valence quarks and antiquarks was independently proposed by Murray Gell-Mann~\cite{Gell-Mann:1964ewy} and George Zweig~\cite{zweig1980developments} in $1964$. This classification was called the quark model, it basically separates hadrons in two big families: mesons and baryons. They are, respectively, quark-antiquark and three-quark bound-states located at the multiplets of the flavour symmetry; all hadrons which belong to the same multiplet have roughly the same mass. The quark model received experimental verification beginning in the late $1960$s and, despite extensive experimental searches, no unambiguous candidates for exotic quark-gluon configurations were identified until the last decade (the interested reader is referred to the Particle Data Group and its topical mini-reviews~\cite{ParticleDataGroup:2020ssz}).

In $2003$, the Belle Collaboration discovered an unexpected enhancement at $3872\,\text{MeV}$ in the $\pi^{+}\pi^{-}J/\psi$ invariant mass spectrum while studying the reaction $B^{+}\to K^{+}\pi^{+}\pi^{-}J/\psi$. The $X(3872)$ state was later studied by the CDF, D0, and BaBar collaborations confirming that its quantum numbers, mass and decay patterns make it an unlikely conventional charm--anti-charm (charmonium) candidate. Therefore, the simple quark model picture that had been so successful for around 35 years was challenged leading to an explosion of related theoretical and experimental activity since then. Nowadays, the number of exotic states has increased dramatically, in both light- and heavy-quark hadron sectors.

The ultimate goal of theory is to describe the properties of exotic states from QCD's first principles. However, since this task is quite challenging, a more modest goal to start with is the development of QCD-motivated phenomenological models that specify the colored constituents, how they are clustered, and the forces between them. This Special Issue invited contributions reporting recent advances of phenomenological quark models in the study of hadron's spectrocopy, structure, and interactions, paying special attention to the exotic candidates but without losing sight of the conventional states. 

\section{Contributions}

After a comprehensive peer review process, 8 articles qualified for acceptance in the final edition of the Special Issue on Advances in Chiral Quark Models. We are now going to summarize the most relevant features of them, encouraging the interested reader to have a look on the contribution itself for further details. 

Chiral symmetry, and its dynamical breaking, has become a cornerstone in the description of the hadron's phenomenology at low energy. Reference~\cite{Fernandez:2021zjq} gives a historical survey on how the quark model of hadrons has been implemented along the last decades trying to incorporate, among other important non-perturbative features of quantum chromodynamics (QCD), the dynamical chiral symmetry breaking mechanism. This effort has delivered different models such as the chiral bag model, the cloudy bag model, the chiral quark model or the chiral constituent quark model. This reference also contributes to the clarification of the differences among the above-mentioned models that include the adjective chiral in their nomenclature.

Reference~\cite{Ortega:2020tng} emphasizes that the understanding of the high energy spectrum of heavy mesons demands the coupling of naive quark-antiquark states with the closest two-meson thresholds. Moreover, this is most likely to be also the case for light quark sectors and even for baryons when considering the meson-baryon thresholds. The review article revisits many examples in which the naive chiral constituent quark model can be complemented with the coupling to two hadron thresholds in order to understand not only the masses but, in some cases, unexpected decay properties of exotic hadron states.

With the development of high energy physics experiments, a large amount of tetraquark and pentaquark candidates in the hadron sector have been observed. In order to shed some insights on the nature of these states, the authors of Ref.~\cite{Yang:2020atz} use a chiral constituent quark model formulated in real- and complex-range approximations in order to study bound, resonance and continuum states of doubly- and all-heavy tetraquarks, but also hidden-charm, hidden-bottom and doubly charmed pentaquarks. Several experimentally observed exotic hadrons fit well in the prediction of such an approach; moreover, their possible compositeness were suggested and other properties analyzed accordingly, \emph{e.g.} decay widths and general patterns in the spectrum. Besides, we report also some predictions not yet seen by experiment within the studied tetraquark and pentaquark sectors.

In Ref.~\cite{Shen:2020gpw}, the authors analyze possible singularities in the $J/\psi \Lambda$ invariant mass distribution of the $\Xi^-_{b}~\to~K^- J/\psi \Lambda$ process via triangle loop diagrams. Triangle singularities in the physical region are found in 18 different triangle loop diagrams. Among those with $\Xi^*$-charmonium-$\Lambda$ intermediate states, the one from the $\chi_{c1} \Xi(2120) \Lambda$ loop, which is located around 4628~MeV, is found the most likely to cause observable effects. One needs $S$- and $P$-waves in $\chi_{c1} \Lambda$ and $J/\psi \Lambda$ systems, respectively, when the quantum numbers of these systems are $1/2^+$ or $3/2^+$. When the quantum numbers of the $\Xi(2120)$ are $J^P=1/2^+$, $1/2^-$ or $3/2^+$, the peak structure should be sharper than the other $J^P$ choices. This suggests that although the whole strength is unknown, one should pay attention to the contributions from the $\Xi^*$-charmonium-$\Lambda$ triangle diagram if structures are observed in the $J/\psi \Lambda$ invariant mass spectrum experimentally. In addition, a few triangle diagrams with the $D_{s1}^*(2700)$ as one of the intermediate particles can also produce singularities in the $J/\psi\Lambda$ distribution, but at higher energies above 4.9~GeV.

Coming back to conventional hadrons but working with a different chiral formalism, the authors of Ref.~\cite{Weigel:2021pbr} review the computations of polarized and unpolarized nucleon structure functions within the bosonized Nambu-Jona-Lasinio chiral soliton model. They focus on a consistent regularization prescription for the Dirac sea contribution and present numerical results from that formulation. Moreover, they also reflect on previous calculations on quark distributions in chiral quark soliton models and attempt to put them into perspective.


All computations mentioned until now are related with low-energy QCD phenomenology at zero temperature and chemical potential. However, the phase diagram of QCD has attracted much attention in the last decades. Beginning with Ref.~\cite{Dumm:2021vop}, the authors review the current status of the research on effective nonlocal NJL-like chiral quark models with separable interactions, focusing on the application of this approach to the description of the properties of hadronic and quark matter under extreme conditions. The analysis includes the predictions for various hadron properties in vacuum, as well as the study of the features of deconfinement and chiral restoration phase transitions for systems at finite temperature and/or density. The authors also address other related subjects, such as the study of phase transitions for imaginary chemical potentials, the possible existence of inhomogeneous phase regions, the presence of color superconductivity, the effects produced by strong external magnetic fields, and the application to the description of compact stellar objects. 

Strong magnetic fields impact quantum-chromodynamics (QCD) properties in several situations; examples include the early universe, magnetars, and heavy-ion collisions. These examples share a common trait-time evolution. A prominent QCD property impacted by a strong magnetic field is the quark condensate, an approximate order parameter of the QCD transition between a high-temperature quark-gluon phase and a low-temperature hadronic phase. The authors of Ref.~\cite{Krein:2021sco} use the linear sigma model with quarks to address the quark condensate time evolution under a strong magnetic field. They use the closed time path formalism of nonequilibrium quantum field theory to integrate out the quarks and obtain a mean-field Langevin equation for the condensate. The Langevin equation features dissipation and noise kernels controlled by a damping coefficient. They compute the damping coefficient for magnetic field and temperature values achieved in peripheral relativistic heavy-ion collisions and solve the Langevin equation for a temperature quench scenario. The magnetic field changes the dissipation and noise pattern by increasing the damping coefficient compared to the zero-field case. An increased damping coefficient increases fluctuations and time scales controlling condensate's short-time evolution, a feature that can impact hadron formation at the QCD transition. The formalism developed here can be extended to include other order parameters, hydrodynamic modes, and system's expansion to address magnetic field effects in complex settings as heavy-ion collisions, the early universe, and magnetars. 

Finally, review~\cite{Yang:2020ykt} introduces the origin of the chiral chemical potential and its physical effects. These include: (1) the chiral imbalance in the presence of strong magnetic and related physical phenomena; (2) the influence of chiral chemical potential on the QCD phase structure; and (3) the effects of chiral chemical potential on quark stars. Moreover, the authors propose for the first time that quark stars are likely to be a natural laboratory for testing the destruction of strong interaction CP.

\section{Conclusions}

QCD is the strong-interaction part of the Standard Model of Particle Physics and solving it presents a fundamental problem that is unique in the history of science. Never before have we been confronted by a theory whose elementary  excitations (quarks and gluons) are not those degrees-of-freedom readily  accessible via experiment; \emph{i.e.}, whose elementary excitations are confined. Moreover, there are numerous reasons to believe that QCD generates forces which are so strong that less-than 2\% of a nucleon's mass can be attributed to the so-called current-quark masses that appear in QCD's Lagrangian; \emph{viz.}, forces capable of generating mass from nothing, a phenomenon known as dynamical chiral symmetry breaking (DCSB). Neither confinement nor DCSB is apparent in QCD's Lagrangian and yet they play 
the dominant role in determining the observable characteristics of hadrons at either zero or finite temperature and chemical potential.

With the completion of this Special Issue, we have mostly described how chiral symmetry can be implemented in the bag model (chiral bag model, little bag model, cloudy bag model) and how the requirement of more flexibility, avoiding the static bag surfaces, gives place to models based on phenomenological Lagrangians such as the Friedberg-Lee model (soliton bag model) and the Nambu--Jona-Lasinio model (chiral quark soliton model). As an improvement of the NJL Hamiltonian, we have introduced the ideas developed by Diakonov based on the instanton liquid model which led to soliton type-solutions of baryons and then becoming to a variant of the chiral quark soliton model; but it also constitutes a foundation of the constituent quark model. A non-relativistic reduction of the Diakonov's Lagrangian gives rise to a chiral constituent quark model which includes gluons and Goldstone-boson exchanges between constituent valence quarks. 

All chiral-based quark models have been applied extensively since their inception to describe hadron phenomenology at zero and finite temperature and baryon density. The authors of this Special Issue have been revisited their own works to provide, first, a general overview of their particular theoretical approach derived from chiral symmetry principles; second, applications of the theoretical formalism to current open problems in hadron physics; and, third, an outlook of the strengths and weaknesses of chiral quark models, possible solutions and future applications.

\vspace{6pt} 

\funding{This work has been partially funded by the Ministerio Espa\~nol de Ciencia e Innovaci\'on under grant No. PID2019-107844GB-C22; and by the Junta de Andaluc\'ia, under contract Nos. P18-FR-5057, Operativo FEDER Andaluc\'ia 2014-2020 UHU-1264517, and PAIDI FQM-370.}

\acknowledgments{The guest editorial team of Symmetry would like to thank all authors for contributing to this Special Issue. The editorial team would also like to thank all anonymous professional reviewers for their valuable time, comments, and suggestions during the review process. We also acknowledge the entire staff of the journal's editorial board for providing their cooperation regarding this Special Issue. We hope that the scientists who are working in the same regime will not only enjoy this Special Issue but also appreciate the efforts made by the entire team.}

\conflictsofinterest{The authors declare no conflict of interest.} 

\end{paracol}
\reftitle{References}


\externalbibliography{yes}
\bibliography{EditorialSpecialIssueSymmetry}

\end{document}